\documentclass[aps,prd,showpacs,preprintnumbers,twocolumn,groupedaddress,nofootinbib]{revtex4-1}

\usepackage[]{amssymb}
\usepackage[]{amsmath}
\usepackage[]{mathrsfs}
\usepackage[]{eucal}
\usepackage[]{tensor}
\usepackage[]{amsthm}
\usepackage[]{bm}

\newcommand{\dd}{\partial}
\newcommand{\df}{\mathrm{d}}
\newcommand{\w}{\wedge}
\newcommand{\Lie}{\pounds}
\newcommand{\nab}[1]{\nabla_{\!#1}}
\newcommand{\heq}[1]{\buildrel #1 \over =}
\newcommand{\qqd}{\ , \quad}
\newcommand{\bc}{\begin{center}}
\newcommand{\ec}{\end{center}}
\newcommand{\be}{\begin{equation}}
\newcommand{\ee}{\end{equation}}

\newcommand{\nn}{\mathbb{N}}

\newcommand{\0}{\emptyset}

\usepackage[unicode]{hyperref}
\usepackage{xcolor}
\definecolor{pastgreen}{HTML}{669900}
\definecolor{pastblue}{HTML}{336699}
\definecolor{linkcol}{HTML}{663333}
\hypersetup{colorlinks,linkcolor={pastblue},citecolor={pastblue},urlcolor={pastblue}}

\theoremstyle{plain} \newtheorem{tm}{Theorem}[section]
\theoremstyle{plain} \newtheorem{lm}[tm]{Lemma}
\theoremstyle{plain} \newtheorem{cor}[tm]{Corollary}
\theoremstyle{definition} \newtheorem{defn}[tm]{Definition}
\theoremstyle{definition} 
\newcommand{\btm}{\begin{tm}}
\newcommand{\etm}{\end{tm}}
\newcommand{\blm}{\begin{lm}}
\newcommand{\elm}{\end{lm}}
\newcommand{\bcor}{\begin{cor}}
\newcommand{\ecor}{\end{cor}}
\newcommand{\bdefn}{\begin{defn}}
\newcommand{\edefn}{\end{defn}}

\begin{document}

\preprint{ZTF-EP-16-04}

\title{Constraints on the symmetry noninheriting scalar black hole hair}

\author{Ivica Smoli\'c}
\email[]{ismolic@phy.hr}

\affiliation{Department of Physics, Faculty of Science, University of Zagreb, P.O.B.~331, HR-10002 Zagreb, Croatia}

\date{\today}

\begin{abstract}
Any recipe to grow black hole hair has to circumvent no-hair theorems by violating some of their assumptions. Recently discovered hairy black hole solutions exist due to the fact that their scalar fields don't inherit the symmetries of the spacetime metric. We present here a general analysis of the constraints which limit the possible forms of such a hair, for both the real and the complex scalar fields. These results can be taken as a novel piece of the black hole uniqueness theorems or simply as a symmetry noninheriting \emph{Ans\"atze} guide. In addition we introduce new classification of the gravitational field equations which might prove useful for various generalizations of the theorems about spacetimes with symmetries.
\end{abstract}


\maketitle

\section{Introduction}

One of the longstanding programs in the black hole physics has been the understanding of the classical fields which can coexist with the black hole horizon. Such fields are generally referred to as a black hole hair and their existence and form have been constrained by a series of no-hair theorems \cite{Heusler,HCC}. Moreover, we now have an opportunity of testing these theorems in the electromagnetic spectrum \cite{Berti15,CHRR15,Test16,J16} and, after the recent detection of gravitational waves by the LIGO Collaboration \cite{LIGO16}, also in the spectrum of gravitational waves \cite{BCW06,BCCC07,GVS12,CFP16,CR16,KZ16}.

\medskip

A typical setting for the analysis of the black hole hair are the spacetimes which admit some symmetries, such as the static and the stationary axisymmetric spacetimes. Arguably the simplest type of hair is grown from the real and the complex scalar fields. The absence of the scalar black hole hair is always proven under some particular assumptions about the scalar field $\phi$, most important being

\begin{itemize}
\item[(a)] a choice of the scalar field coupling to gravity (the way it appears in the matter-gravity field equations),
\item[(b)] an energy condition (e.g.~some inequality imposed on the scalar field potential),
\item[(c)] the assumption that scalar field $\phi$ is nonsingular on the black hole horizon and
\item[(d)] the assumption that scalar field $\phi$ \emph{inherits} the spacetime symmetries: for any Killing vector field $\xi^a$, such that $\Lie_\xi g_{ab} = 0$, we necessarily have $\Lie_\xi \phi = 0$.
\end{itemize}

Bekenstein's original theorems \cite{Beken72,BekenStaticHair,BekenStationaryHair} rest upon the simplest set of choices among these: minimally coupled, canonical scalar fields, regular on the black hole horizon and inheriting all the spacetime symmetries. These theorems were later generalized to a great extent, for the minimally coupled \cite{Beken95,Sud95,BL07,BL11,GJ14a} as well as for the nonminimally coupled \cite{XZ91,Zann95,Saa96a,Saa96b,SZ98,AB02,Winst05,SS14} scalar fields. On the other hand, over the past several decades, a number of hairy black hole solutions have been found \cite{BBM,Beken74,MTZ04,SZ14}, each violating some of the basic assumptions of the no-hair theorems. However, only recently the rotating black holes with the \emph{symmetry noninheriting} complex scalar hair have been found \cite{HR14,HR15}. This begs a question: What are the general symmetry inheritance properties of the scalar fields and in which possible forms can the symmetry noninheriting (sni) scalar fields appear as a black hole hair?

\medskip

After the pioneer paper by Hoenselaers \cite{Hoen78}, dealing with the simplest form of the real scalar fields, and several recent ones \cite{BST10a,BST10b,GJ14b} focused on the more specific time dependence of the real scalar field in a stationary spacetime, the first general analysis of the symmetry inheritance properties of both real and complex scalar fields was presented in \cite{ISm15}. As a direct consequence it was noticed that, given that the spacetime is a solution to the Einstein's gravitational field equations, the assumption about the symmetry inheritance of the canonical real scalar field in the original Bekenstein's theorem \cite{Beken72} is in fact superfluous. As most of the no-hair theorems \cite{Heusler,HCC} and the symmetry inheritance properties proven in \cite{ISm15} are valid for a much general class of the gravitational theories, one might wonder if these two groups of results can be unified in some more effective way. The objective of this paper, thus, is to expand the encyclopaedia of no-hair theorems with a broad treatment of the symmetry noninheriting scalar fields.

\medskip 

In Sec.~II we precisely define the classes of spacetimes with symmetries which are general enough to include many known important solutions, but still useful enough to draw the conclusions that we aspire to. In Sec.~III we single out one crucial tensorial property and prove that some important classes of gravitational field equations do in fact fulfill it. Sections V and VI treat, respectively, the real and the complex scalar fields, proving the central results of the paper. In the final section we summarize the results and raise some open questions.

\subsection{Basic conventions}

We shall use the ``mostly plus'' metric signature $(-,+,+,\dots)$ and the natural system of units, $G = c = 1$. Unless stated otherwise, the total number of spacetime dimensions is always general $D \ge 3$. Tensors are written either in the abstract index notation (see e.g.~\cite{Wald}), denoted by the lowercase Latin letters from the beginning of the alphabet, or in the ``indexless'' notation (see e.g.~\cite{Heusler}), where appropriate. For any symmetric tensor field $S_{ab}$ and a vector field $X^a$ we introduce the 1-form $S(X)_a \equiv S_{ab} X^b$, abbreviated in the indexless notation as $S(X)$.

\medskip

We shall consider the gravitational equations of motion which can be written in a form
\be\label{eq:EOM}
E_{ab} = 8\pi T_{ab} \ ,
\ee
with the energy-momentum tensor $T_{ab}$. The tensor $E_{ab}$ is a general diff-covariant smooth function of the spacetime metric, the Riemann tensor, the Levi-Civita tensor and the covariant derivatives. As a consequence of these assumptions we know that the existence of a Killing vector field $\xi^a$, such that $\Lie_\xi g_{ab} = 0$, implies $\Lie_\xi E_{ab} = 0$ and immediately, via (\ref{eq:EOM}),
\be\label{eq:LieT}
\Lie_\xi T_{ab} = 0 \ ,
\ee
the central equation in the analysis of the symmetry inheritance. We always assume that any scalar field is at least of differentiability class $C^2$.

\medskip

\section{Geometric setting}

We shall consider a spacetime which admits a family of $n \in \nn$ smooth pairwise commuting Killing vector fields $\{\xi^a_{(1)},\dots,\xi^a_{(n)}\}$, everywhere orthogonal to a family of submanifolds which foliate the spacetime. Technically, we demand that the tangent distribution $\Delta^\bot$, orthogonal to the distribution $\Delta$ spanned by these Killing vector fields, must be totally integrable. Frobenius' theorem (see e.g.~\cite{Lee}, chapter 19) is telling us that a tangent distribution is totally integrable if and only if it is involute. The dual form of this theorem states that $\Delta^\bot$ is involute if and only if the 1-forms $\{\xi_a^{(1)},\dots,\xi_a^{(n)}\}$, associated to the Killing vectors that span the $\Delta$, satisfy equations
\be\label{eq:ot}
\xi^{(1)} \w \dots \w \xi^{(n)} \w \df \xi^{(i)} = 0
\ee
for all $i \in \{1,\dots,n\}$. 

\medskip

Just to give some more concrete examples, let us look at the cases of the time-independent and the axially symmetric spacetimes. More precisely, we say that the asymptotically flat spacetime $(M,g_{ab})$ is \emph{stationary} if it allows a complete Killing vector field $k^a$, which is timelike ``at infinity'' (say, in a neighbourhood of the null infinities $\mathscr{I}^\pm$). Furthermore, a spacetime is \emph{axisymmetric} if it allows a spacelike Killing vector field $m^a$ with compact orbits. Two frequently studied subcases of these are 

\begin{itemize}
\item[(i)] \emph{Static spacetime} (e.g.~the Schwarzschild spacetime), which is stationary spacetime, such that 1-form $k_a$, associated to the stationary Killing vector field $k^a$, satisfies
\be
k \w \df k = 0 \ .
\ee 

\item[(ii)] \emph{Circular spacetime} (e.g.~the Kerr spacetime), which is a stationary axi\-symmetric spacetime, such that $[k,m]^a = 0$ and
\be
k \w m \w \df k = k \w m \w \df m = 0 \ .
\ee
\end{itemize}

\medskip

More generally, we have the following definition.

\medskip

\bdefn\label{def:otKill}
We say that a smooth $D$-dimensional spacetime $(M,g_{ab})$ admits $n < D$ \emph{orthogonally-transitive} Killing vectors if it admits $n$ smooth pairwise commuting, (almost everywhere) linearly independent Killing vector fields $\{\xi^a_{(1)},\dots,\xi^a_{(n)}\}$, satisfying conditions (\ref{eq:ot}).
\edefn

\medskip

Commuting of the Killing vectors in Definition \ref{def:otKill} implies that the distribution $\Delta$ is involute, thus integrable with the corresponding family of $n$-dimensional submanifolds (``surfaces of transitivity''), which we shall generically denote by $N \subseteq M$. Furthermore, conditions (\ref{eq:ot}) imply the existence of a family of $(D-n)$-dimensional smooth submanifolds (to which these Killing vector fields are orthogonal), which we shall generically denote by $\Sigma \subseteq M$.

\medskip

\emph{Degenerate points}. Now, building upon this geometric setting we would like to introduce a useful coordinate system, but first we must take special care of the ``degenerate points'', the importance of which will be clarified below. Using an auxiliary function,
\be
W \equiv n!\,\xi^{(1)}_{[a_1} \cdots \xi^{(n)}_{a_n]} \, \xi^{[a_1}_{(1)} \cdots \xi^{a_n]}_{(n)} \ ,
\ee
we define its set of zeros,
\be
\mathscr{W} \equiv \left\{ p \in M \ | \ W(p) = 0 \right\} \ .
\ee
Furthermore, a spacetime may contain a nonempty (albeit usually a zero measure) set of points,
\be
\mathscr{Z} \equiv \left\{ p \in M \ | \ \xi^{(1)} \w \dots \w \xi^{(n)} \heq{p} 0 \right\} \subseteq \mathscr{W}
\ee
at which the condition of linear independence of the Killing vector fields breaks. An example of such a set is the axis of symmetry, where an axial Killing vector field vanishes. 

\medskip

So, in what sense are these points ``degenerate''? Noting that $W = \det(g(\xi_{(i)},\xi_{(j)}))$, it becomes clear that the vanishing of this function is sufficient and necessary condition for the existence of the null Killing vector field which is a linear combination of the Killing vector fields from Definition \ref{def:otKill}, and which is orthogonal to all of them. In other words, the surface of transitivity $N_p$ through a point $p \in M - \mathscr{Z}$ is null (there is a vector field both tangent and orthogonal to $N_p$) if and only if $W(p) = 0$ \cite{Carter69}. A particularly interesting subset of $\mathscr{W}$ are the black hole horizons.

\medskip

\emph{Killing horizons}. There is a multitude of related, but subtly different definitions of the black hole horizons, among which the spacetimes with symmetries allow a particularly useful one. The Killing horizon $H[\xi]$, associated to the Killing vector field $\xi^a$, is a null embedded (not necessarily connected) hypersurface, invariant under the flow of $\xi^a$ which is null and nonvanishing on $H[\xi]$. The closure of a Killing horizon may fail to be a submanifold if its components ``cross'' in a spacelike codimension-2 bifurcation surface $\mathcal{B}[\xi]$ on which $\xi^a$ vanishes, so that such an union of subsets is usually referred to as a bifurcate Killing horizon. As most of our arguments on the black hole horizon will be local, formal difference between the ``ordinary'' and the ``bifurcate'' Killing horizons is not relevant here. If the Killing vector field $\xi^a$ is a linear combination of the Killing vector fields from Definition \ref{def:otKill} then $H[\xi] \subseteq \mathscr{W}$ and $\mathcal{B}[\xi] \subseteq \mathscr{Z}$. For later convenience we denote the Killing horizon with excluded points from the set $\mathscr{Z}$ (bifurcation surface, intersection of the axis of symmetry with the horizon) by
\be
H[\xi]^\times \equiv H[\xi] \cap (M - \mathscr{Z})
\ee
or just shortly $H^\times$. Typical exact stationary black hole solutions, such as the ones from the Kerr-Newman family, contain a Killing horizon. A black hole binary is not longer a stationary system, but still can be modeled by a spacetime possessing so-called helical Killing vector field \cite{FUS02} and the associated Killing horizon. Finally, we note in passing that the cosmological horizons also provide another example of the Killing horizons.

\medskip

As the Killing vector field are by assumption smooth, the function $W$ is at least continuous and the set $\mathscr{W}$ is thus closed. We shall consider spacetimes in which the degenerate points appear only on the boundary of the ``regular regions'', thus we always assume that the interior $\mathscr{W}^\circ$ is empty, which implies that $\dd\mathscr{W} = \mathscr{W}$. Most of the constructions and proofs that follow will be performed on the open set $M - \mathscr{W}$, with conclusions extended by continuity of the various tensor fields to the boundary $\dd\mathscr{W}$. In this respect we are relying on the same strategy that was used in \cite{CDPS15} (encapsulated in Lemma 2 and Lemma 3 therein). In addition, we emphasize that, assuming that $M$ is a connected manifold and $\mathscr{W} \ne \0$, any connected component of the set $M - \mathscr{W}$ is an open set $O$ with a nonempty boundary (otherwise the set $O$ would be both open and closed in $M$, in contradiction with the connectedness of $M$) $\dd O \subseteq \mathscr{W}$.

\medskip

As a consequence of the assumptions given in Definition \ref{def:otKill} and all the remarks from above, at each point $p \in M - \mathscr{W}$ the tangent space splits as $T_p M = T_p N \oplus T_p \Sigma$. This allows us to choose a (local) coordinate system 
\be\label{coord}
\{z^1,\dots,z^n,y^1,\dots,y^{D-n}\} \ ,
\ee
such that
$$\xi^a_{(i)} = \left( \frac{\dd}{\dd z^i} \right)^{\!a} \in T_p N \quad \textrm{and} \quad Y^a_{(A)} \equiv \left( \frac{\dd}{\dd y^A} \right)^{\!a} \in T_p \Sigma \ ,$$
where $i \in \{1,\dots,n\}$, $A \in \{1,\dots,D-n\}$ and
\be
[\xi_{(i)},Y_{(A)}]^a = 0 \qqd g_{iA} = g(\xi_{(i)},Y_{(A)}) = 0
\ee 
for any $i$ and $A$. In order to emphasize that tensor indices take value in some of these two subsets of coordinates, we shall call them ``$z$-block'' and ``$y$-block'' indices. In the coordinate system (\ref{coord}), the spacetime metric has the ``block-diagonal form'' \cite{Carter69}
\be\label{eq:blockm}
\df s^2 = g_{ij}(y)\,\df z^i \, \df z^j + g_{AB}(y)\,\df y^A \, \df y^B \ .
\ee
Apart from the members of the Kerr-Newman family of spacetimes, nontrivial examples admitting orthogonally-transitive Killing vector fields include the Myers-Perry black hole spacetime \cite{MP86} and the generalized Weyl solutions \cite{ER02}. We collect a useful set of properties of the metric (\ref{eq:blockm}) in the following Lemma.

\medskip

\blm\label{lm:comp}
Let $(M,g_{ab})$ be a spacetime with the metric of the form (\ref{eq:blockm}). Then, in the coordinate system (\ref{coord}) introduced above, the vanishing components of the Christoffel symbol are
\be
\Gamma^i_{jk} = \Gamma^i_{AB} = \Gamma^A_{iB} = 0 \ ,
\ee
while the vanishing components of the Riemann tensor and its covariant derivatives are
\be
R_{iABC} = R_{Aijk} = 0 \qqd R_{iA} = 0 \ ,
\ee
$$\nab{i} R_{jklm} = \nab{i} R_{jAkB} = \nab{i} R_{jkAB} =$$
\be
= \nab{i} R_{ABCD} = \nab{A} R_{Bijk} = \nab{A} R_{BCDi} = 0 \ .
\ee
For any smooth real function $f$, such that $\dd_i f = 0$ for all $i \in \{1, \dots, n\}$, we have
\be
\nab{i}\nab{A} f = \dd_i \dd_A f - \Gamma^k_{iA} \dd_k f = 0 \ .
\ee 
\elm

\medskip

\section{Assorting the gravitational field equations}

An important property of the Ricci tensor, that has been widely used in various uniqueness theorems \cite{Heusler}, is that the static metric is necessarily \emph{Ricci static}, 
\be 
k \w R(k) = 0 \ ,
\ee
and that the circular metric is necessarily \emph{Ricci circular}, 
\be
k \w m \w R(k) = 0 = k \w m \w R(m) \ .
\ee
Since our goal is to extend the conclusions to the gravitational field equations as general as possible, we introduce the definition that provides a natural extension of the aforementioned terms.

\medskip

\bdefn\label{def:otE}
Let $(M,g_{ab})$ be a smooth $D$-di\-men\-si\-onal spacetime allowing $n < D$ orthogonally-transitive Killing vector fields $\{\xi^a_{(1)},\dots,\xi^a_{(n)}\}$, such that the set of degenerate points has empty interior, $\mathscr{W}^\circ = \0$. Then we say that a tensor $E_{ab}$ is a member of the \emph{orthogonal-transitive class} (``o-t class'') \emph{of order} $n$ if the conditions
\be\label{eq:Eot}
\xi^{(1)} \w \dots \w \xi^{(n)} \w E(\xi_{(i)}) = 0
\ee
hold for all $i \in \{1,\dots,n\}$.
\edefn

\medskip

In other words, the condition (\ref{eq:Eot}) holds if and only if 
\be\label{eq:EiA}
E_{iA} = E(\xi_{(i)},Y_{(A)}) = 0
\ee
holds in the coordinate system (\ref{coord}). In practice we need to check the condition (\ref{eq:EiA}) on the set $M - \mathscr{W}$ and then, using the continuity of all the tensors involved here, extend the validity of (\ref{eq:Eot}) to the degenerate points $\mathscr{W}$ on the boundary. Of course, this generalization of the Ricci staticity and the Ricci circularity would be useless if the physically motivated tensors $E_{ab}$ belonging to some o-t class are rare. The situation, as we shall immediately see, is quite the opposite. Using the results in Lemma \ref{lm:comp} it can be easily checked that the following two examples of tensors satisfy the property (\ref{eq:Eot}):

\medskip

\textbf{(i) GR}. The ``gravitational side'' of the Einstein's field equation is the sum of the Einstein's tensor $G_{ab}$ and the cosmological constant term,
\be\label{eq:Einstein}
E_{ab}^{(E)} = R_{ab} - \frac{1}{2}\,R g_{ab} + \Lambda g_{ab} \ .
\ee

\medskip

\textbf{(ii)} $\mathbf{f(R)}$. Based on theoretical models of quantum gravity and phenomenological problems in cosmology and astrophysics, various modifications of the general relativity have been proposed. Among these, extensively studied family of $f(R)$ theories \cite{SF10,DeFT10} is simply constructed by replacement of the Einstein-Hilbert lagrangian term $R$ with some general differentiable function $f(R)$. Correspondingly, the Einstein tensor generalizes to
\begin{align}\label{eq:fR}
E_{ab}^{(f)} & = f'(R) R_{ab} - \frac{1}{2}\,f(R)g_{ab} - \nonumber\\
 & - \left( \nab{a}\nab{b} - g_{ab}\,\Box \right) f'(R) \ .
\end{align}

\medskip

A less trivial class of examples appears when multiple Riemann tensors are contracted in the gravitational equation of motion.

\medskip

\textbf{(iii) Lovelock gravity}. Lovelock has found \cite{Love71} the natural extension of the Einstein's tensor to higher dimensions: The most general rank-two symmetric divergence-free tensor, built from the spacetime metric and its first and second derivatives, is given by
\begin{align}\label{eq:Love}
E_{ab}^{(L)} & = \lambda_0 g_{ab} + \sum_{p=1}^{\lfloor \frac{D-1}{2} \rfloor} \lambda_p \, g_{ac} \, \delta^{c d_1 \dots d_{2p}}_{b e_1 \dots e_{2p}} \ \times \nonumber\\
 & \times \tensor{R}{^{e_1}^{e_2}_{d_1}_{d_2}} \cdots \tensor{R}{^{e_{2p-1}}^{e_{2p}}_{d_{2p-1}}_{d_{2p}}}
\end{align}
with some arbitrary real constants $\{\lambda_0,\lambda_1,\dots\}$. Let us now examine more closely the $E_{iA}$ component,
\begin{align}\label{eq:LoveiA}
E_{iA}^{(L)} & = \sum_{p=1}^{\lfloor \frac{D-1}{2} \rfloor} \lambda_p \, g_{ij} \, \delta^{j \mu_1 \dots \mu_{2p}}_{A \sigma_1 \dots \sigma_{2p}} \ \times \nonumber\\
 & \times \tensor{R}{^{\sigma_1}^{\sigma_2}_{\mu_1}_{\mu_2}} \cdots \tensor{R}{^{\sigma_{2p-1}}^{\sigma_{2p}}_{\mu_{2p-1}}_{\mu_{2p}}} \ .
\end{align}
Let $\ell_1$ be the number of ``$z$-block'' indices among the ``$\mu$ indices'', and $\ell_2$ the number of ``$z$-block'' indices among the ``$\sigma$ indices'' in the generalized Kronecker delta appearing in the sum (\ref{eq:LoveiA}). As any nonvanishing component of the Riemann tensor contains even number of indices from any block (either all indices from one block, or we have two and two from each), the difference $\ell_1 - \ell_2$ in the nonvanishing terms of the sum (\ref{eq:LoveiA}) must be an even integer. On the other hand, the total number of contravariant ``$z$-block'' indices and the total number of covariant ``$z$-block'' indices in the nonvanishing components of the generalized Kronecker delta should be equal. As the generalized Kronecker delta contains one extra contravariant ``$z$-block'' index (namely, ``$j$''), this is impossible and all terms in the sum (\ref{eq:LoveiA}) vanish. Thus, we may conclude that in fact
\be
E^{(L)}_{iA} = 0 \ , 
\ee
so that any tensor of the form (\ref{eq:Love}) is a member of the o-t class of any order $n \in \nn$.

\medskip

\emph{Nonexamples}. On the other hand, there is a significant family of tensors which in general do not belong to any o-t class. Gravitational field equations in odd $D = 2m - 1$ dimensional spacetimes may contain gravitational Chern-Simons terms \cite{DJT82,Witten88,BCPPS11a},
\begin{align}\label{eq:Cotton}
C^{ab} & = -\frac{m}{2^{m-1}}\,\epsilon^{c_1 \dots c_{2m-2} (a} \nab{e} \Big( \tensor{R}{^{b)}_{d_1}_{c_1}_{c_2}} \tensor{R}{^{d_1}_{d_2}_{c_3}_{c_4}} \cdots \nonumber\\
 & \cdots \tensor{R}{^{d_{m-3}}_{d_{m-2}}_{c_{2m-5}}_{c_{2m-4}}} \tensor{R}{^{d_{m-2}}^e_{c_{2m-3}}_{c_{2m-2}}} \Big)
\end{align}
motivated by the low-energy superstring effective actions. Although the generalized Cotton tensor (\ref{eq:Cotton}) identically vanishes for a class of block-diagonal metrics (see theorem 1 in \cite{BCDPPS13}), our metric (\ref{eq:blockm}) generally cannot be written in this form. For example, the 2-form $k \w C(k)$ doesn't necessarily vanish for a static metric with the associated stationary Killing vector field $k^a$, a fact which was noticed for the 3-dimensional case in \cite{AN96}. It would be interesting to see whether the generalized Cotton tensor belongs to some of the o-t classes under certain additional assumptions. 

\medskip

For later convenience we introduce one practical abbreviation.

\medskip

\bdefn
We say that the spacetime $(M,g_{ab},\psi)$ (with some matter field $\psi$) is an \emph{o-t symmetric solution} of (\ref{eq:EOM}) \emph{of order $n$} if the tensor $E_{ab}$ is a member of the o-t class of order $n$ (as defined in \ref{def:otE}) and if the spacetime allows a set of $n$ hypersurface orthogonal Killing vector fields $\{\xi^a_{(1)},\dots,\xi^a_{(n)}\}$ with the set of degenerate points $\mathscr{W}$ which has empty interior. 
\edefn

\medskip

Now, let us assume that the spacetime from Definition \ref{def:otKill} is a solution of the field equation (\ref{eq:EOM}), where $E_{ab}$ is a member of a o-t class of the order $n$. Then the same is true also for the energy-momentum tensor $T_{ab}$,
\be\label{eq:Tot}
\xi^{(1)} \w \dots \w \xi^{(n)} \w T(\xi_{(i)}) = 0 \ .
\ee
We note in passing that in the special cases of the static and the circular spacetimes, corresponding properties of the energy-momentum tensor, represented by the relation (\ref{eq:Tot}), are usually referred to as the ``matter staticity'' and the ``matter circularity'' \cite{Heusler}.

\medskip

\section{Spacetimes with black holes}

We can now add the final part of our geometric setting for the spacetimes with symmetries, a black hole.

\medskip

\bdefn\label{def:assKH}
Let $(M,g_{ab})$ be a spacetime admitting a set of $n$ orthogonally-transitive Killing vector fields $\{\xi^a_{(1)},\dots,\xi^a_{(n)}\}$. Then we say that a Killing horizon $H[\chi]$ is \emph{associated} to these Killing vector fields if 

\begin{itemize}
\item[(i)] $H[\chi]$ is generated by the Killing vector field
\be\label{eq:chi}
\chi^a = \sum_{i=1}^n b_i\,\xi^a_{(i)} \ ,
\ee
defined with some real constants $\{b_1,\dots,b_n\}$ which are not all zero,

\item[(ii)] $H[\chi]$ is invariant under the flow of the Killing vector fields $\xi^a_{(i)}$ for all $i \in \{1,\dots,n\}$ and

\item[(iii)] $H[\chi]$ intersects the boundary of each connected component of $M - \mathscr{W}$.
\end{itemize} 
\edefn

\medskip

From the definition above, it follows that any Killing vector field $\xi^a_{(i)}$ is tangent to the associated Killing horizon $H[\chi]$, thus $\chi_a \xi_{(i)}^a = 0$ holds on $H[\chi]$ for all $i$. Then (\ref{eq:Eot}) implies that
\be
\xi^{(1)} \w \dots \w \xi^{(n)} \w E(\chi) = 0 \ ,
\ee
which contracted with $\chi^a$ on $H[\chi]^\times$ implies that
\be\label{eq:Echichi}
E(\chi,\chi) \heq{H^\times} 0 \ .
\ee
The field equation (\ref{eq:EOM}) implies that the energy-momentum tensor $T_{ab}$ satisfies the analogous relation
\be\label{eq:Tchichi}
T(\chi,\chi) \heq{H^\times} 0 \ .
\ee
This equation represents a valuable boundary condition which we shall exploit for the central results in this paper. Relations of the form (\ref{eq:Echichi}) and (\ref{eq:Tchichi}) are well known in the basic case of the Einstein's gravitational field equations, and have been recently proven for the spacetimes which fall into some of the o-t classes considered here and which are solutions of the Lovelock's \cite{SB13} and the $f(R)$ \cite{Bhatt16} gravitational field equations. The importance of these relations stems from their roles in various black hole theorems, such as the zeroth law of black hole thermodynamics \cite{Wald} and the constancy of the electric and the magnetic potentials on the Killing horizons \cite{ISm12,ISm14}.

\medskip

The most interesting cases for us are those of the static spacetimes (with associated Killing vector field $k^a$), containing a nonrotating Killing horizon $H[k]$, and the stationary axisymmetric spacetimes (with associated ``stationary'' Killing vector field $k^a$ and $\ell \in \nn$ spacelike ``axial'' Killing vector fields $m^a_{(i)}$ with compact orbits), containing a Killing horizon $H[\chi]$, generated by the Killing vector field
\be\label{eq:chikm}
\chi^a = k^a + \sum_{i=1}^\ell \Omega_i m_{(i)}^a \ .
\ee
The real constants $\Omega_i$ are usually referred to as the ``angular momenta'' of the Killing horizon $H[\chi]$.

\medskip

\section{Real Scalar Fields}

The simplest matter content of a spacetime is provided by the real scalar field. Here we shall consider only minimally coupled real scalar fields, first those described by the ``canonical'' energy-momentum tensor, then a larger class of ``noncanonical'' ones.

\medskip

\emph{Canonical case}. The canonical real scalar field is described with the energy-momentum tensor of the form 
\be\label{eq:canonT}
T_{ab} = \nab{a}\phi \nab{b}\phi + \left( X - V(\phi) \right) g_{ab} \ ,
\ee
where we have introduced the usual abbreviation
\be
X = -\frac{1}{2}\,\nab{c}\phi \nabla^c \phi
\ee
and $V(\phi)$ is some potential. From the theorems in \cite{ISm15} we know that the only possible form of the noninheriting scalar field appears in the case when the Killing vector field $\xi^a$ has noncompact orbits and the potential $V(\phi)$ is constant. Furthermore, the aforementioned theorem asserts that the Lie derivative $\Lie_\xi \phi$ must be constant on each connected component of $M - \mathscr{W}$ (thus, the field $\phi$ is unbounded along the complete noncompact orbits of $\xi^a$). For example, an explicit example of a stationary spacetime with the massless real scalar field linearly growing with time was found by Wyman \cite{Wyman81}. In a spacetime with a set of $n$ hypersurface-orthogonal Killing vector fields and the associated Killing horizon $H[\chi]$, the relation (\ref{eq:Tchichi}) implies that
\be
\Lie_\chi \phi \heq{H^\times} 0 \ .
\ee
Supposing that, without loss of generality, the first $r \le n$ among the Killing vector fields have noncompact orbits, then this relation provides us with an useful constraint, as summarized in the following theorem.

\medskip

\btm
Let the spacetime $(M,g_{ab},\phi)$ be an o-t symmetric solution of (\ref{eq:EOM}) of order $n$ with the energy-momentum tensor (\ref{eq:canonT}), containing the associated Killing horizon $H[\chi]$ (as defined in \ref{def:assKH}). If the first $r \le n$ among the Killing vector fields have noncompact orbits, then the following relation holds
\be
\sum_{i=1}^r b_i \Lie_{\xi_{(i)}} \phi = 0 \ ,
\ee
where each term $\Lie_{\xi_{(i)}} \phi$ is a constant. 
\etm

\medskip

If only one Killing vector field has noncompact orbit ($r = 1$), with nonvanishing constant $b_1 \ne 0$, then $\phi$ necessarily inherits all the symmetries. For example, this is the case that we encounter in the static spacetimes with the nonrotating Killing horizon $H[k]$, or the stationary axisymmetric spacetimes with the corresponding Killing horizon $H[\chi]$. Therefore, the assumption about the symmetry inheritance of the real scalar fields in the original Bekenstein's no-hair theorem \cite{Beken72} is in fact superfluous, at least when the tensor $E_{ab}$ is a member of the o-t class of the order 1 (in the static case) or of the order 2 (in the circular case).

\bigskip

\emph{Noncanonical case}. The ``$k$-essence'' theories \cite{APDM99,APMS01} generalize the form of the scalar field energy-momentum tensor,
\be\label{eq:noncT}
T_{ab} = p_{,X} \nab{a}\phi \nab{b}\phi + p g_{ab}
\ee
with some function $p = p(\phi,X)$. The canonical case is recovered with the special choice $p(\phi,X) = X - V(\phi)$. From the results in \cite{ISm15} we know that $Xp_{,X}$ is constant along the orbits of $\xi^a$, $\Lie_\xi (Xp_{,X}) = 0$, which allows us to construct one classification of the symmetry inheritance properties. 

\medskip

Let $\gamma$ be a complete orbit of $\xi^a$ which doesn't contain ``exceptional points'', those in which all three functions $p_{,\phi}$, $p_{,X}$ and $\Lie_\xi p_{,X}$ vanish. The theorem 4 in \cite{ISm15} then distinguishes following two cases

\begin{itemize}
\item[(a)] $Xp_{,X} = 0$ on $\gamma$ implies that $\Lie_\xi \Lie_\xi \phi = 0$ holds along $\gamma$ for an admissible\footnote{We say that the energy momentum $T_{ab}$ is \emph{admissible} if the function $p$ is such that $\Lie_v(p_{,X})$ vanishes for an arbitrary vector field $v^a$ whenever both $X$ and $\Lie_v X$ vanish} energy-momentum tensor $T_{ab}$;

\item[(b)] $Xp_{,X} \ne 0$ on $\gamma$ implies that $\Lie_\xi \phi$ is a solution to the differential equation
\be\label{eq:Y}
p_{,\phi} (\Lie_\xi \phi)^2 + 2Xp_{,X} \Lie_\xi (\Lie_\xi \phi) = 0 \ ,
\ee
which is either a positive definite function, a negative definite function or identically zero on $\gamma$.
\end{itemize}

\medskip

What are the implications of this theorem on the symmetry inheritance? Obviously, none of these cases is reconcilable with a compact orbit $\gamma$ unless $\Lie_\xi \phi = 0$ along $\gamma$. A noncompact orbit requires some additional elaboration. In the case (a) the field $\phi$ either inherits symmetry or is unbounded along $\gamma$ (a case which might be disregarded as unphysical). Let us focus on the case (b) and assume, without loss of generality, that $\Lie_\xi \phi > 0$ holds along $\gamma$. If $\phi$ remains bounded on $\gamma$ then $\Lie_\xi \phi$ has to obtain arbitrarily small values in both directions along this orbit (if the limit of $\Lie_\xi \phi$ along the orbit exists then it must be zero). But then $\Lie_\xi \phi$ has to have local extrema, a point $s \in \gamma$ where $\Lie_\xi \Lie_\xi \phi = 0$. Equation (\ref{eq:Y}) here implies that either $\Lie_\xi \phi = 0$ (in contradiction with the assumption $\Lie_\xi \phi > 0$) or $p_{,\phi} = 0$ at point $s$. Thus, if $Xp_{,X} \ne 0$ along a noncompact orbit $\gamma$ which doesn't contain ``exceptional points'' nor the points on which both $p_{,\phi}$ and $\Lie_\xi \Lie_\xi \phi$ vanish, then the scalar field $\phi$ either inherits symmetry or is unbounded along $\gamma$.

\bigskip

Now, using the geometric setting introduced above, we can make some further conclusions about the noncanonical scalar fields. By inserting (\ref{eq:EiA}) and (\ref{eq:noncT}) into (\ref{eq:EOM}) we get that the relation
\be\label{eq:pxiY}
p_{,X} \left( \Lie_{\xi_{(i)}} \phi \right) \left( \Lie_{Y_{(A)}} \phi \right) = 0
\ee
holds on $M - \mathscr{W}$ for each $\xi_{(i)}^a$ and $Y_{(A)}^a$. Whence, at each point $q \in M - \mathscr{W}$ where $p_{,X} \ne 0$ holds and for at least one Killing vector field we have symmetry noninheritance, $\Lie_{\xi_{(i)}} \phi \ne 0$, the relation (\ref{eq:pxiY}) implies that $\Lie_{Y_{(A)}} \phi = 0$ holds for all $Y_{(A)}^a \in T_q \Sigma$.

\medskip

Just to be little more concrete, let us look more closely at the case when only one Killing vector field has noncompact orbits, $k^a = (\dd/\dd t)^a$. Using the fact that $\phi$ inherits all the symmetries generated by the Killing vector fields with compact orbits, and consequences of Eq.~(\ref{eq:pxiY}), we know that $\phi = \phi(t)$. If the spacetime contains a bifurcate Killing horizon $H[\chi]$, generated by the Killing vector field of the form (\ref{eq:chikm}), then $\Lie_k \phi$ vanishes on the bifurcation surface $\mathcal{B}[\chi]$. Furthermore, suppose that surfaces of constant $t$ intersect the bifurcation surface $\mathcal{B}[\chi]$ (see e.g.~Sec.~5 in \cite{RW92}), which is a subset of the open set where $X p_{,X} \ne 0$. Then, as $\Lie_k \phi$ is by assumption a continuous function, we may deduce that $\Lie_k \phi$ vanishes on each surface of constant $t$ and that $\phi$ inherits all the symmetries.

\medskip

\section{Complex Scalar Fields}

The canonical energy-momentum tensor for the complex scalar field $\phi$ is given by
\be\label{eq:Tcmplx}
T_{ab} = \nab{(a}\phi \nab{b)}\phi^* - \frac{1}{2} \, \big( \nab{c}\phi \nabla^c\phi^* + V(\phi^*\phi) \big) g_{ab} \ .
\ee
We shall use the polar form of the complex scalar field,
\be
\phi = A\,e^{i\alpha}
\ee
where $A$ is the amplitude and $\alpha$ is the phase of the field $\phi$. In order to leave the trivial case aside, in the rest of the chapter we shall focus only on the points from the open set $Q \subseteq M - \mathscr{W}$ where $A \ne 0$. In this parametrization, the energy momentum tensor (\ref{eq:Tcmplx}) becomes
\be
T_{ab} = \nab{a}A \nab{b}A + A^2 \nab{a}\alpha \nab{b}\alpha + \frac{T+V(A^2)}{D-2}\,g_{ab} \ ,
\ee
where $T \equiv g^{ab} T_{ab}$. In order to simplify the notation we shall use the abbreviations
$$\dot{f} = \Lie_{\xi_{(i)}} f \quad \textrm{and} \quad f' = \Lie_{Y_{(A)}} f$$
for any function $f$.
From Eqs.~(\ref{eq:EOM}) and (\ref{eq:EiA}), we have 
\be\label{eq:TxiY}
0 = T(\xi_{(i)},Y_{(A)}) = \dot{A} A' + A^2 \dot{\alpha} \alpha' \ .
\ee
It is not too difficult to see that this system of equations drives the coordinate dependence ``synchronisation'' of the amplitude $A$ and the phase $\alpha$. If $p \in Q$ and there is

\begin{itemize}
\item[(a)] $Y^a_{(A)} \in T_p \Sigma$ such that $A' \ne 0$ and $\alpha' = 0$, then $\dot{A} = 0$ for all $\xi^a_{(i)} \in T_p N$. In addition, assuming that there is $\xi^a_{(i)}$, such that $\dot{\alpha} \ne 0$ (otherwise, we would have the complete symmetry inheritance), we can conclude that $\alpha' = 0$ for all $Y^a_{(A)} \in T_p \Sigma$.

\item[(b)] $Y^a_{(A)} \in T_p \Sigma$ such that $A' = 0$ and $\alpha' \ne 0$ then $\dot{\alpha} = 0$ for all $\xi^a_{(i)} \in T_p N$. In addition, assuming that there is $\xi^a_{(i)} \in T_p N$, such that $\dot{A} \ne 0$ (otherwise we would have the complete symmetry inheritance), we can conclude that $A' = 0$ for all $Y^a_{(A)} \in T_p \Sigma$.

\item[(c)] $\xi^a_{(i)} \in T_p N$ such that $\dot{A} = 0$ and $\dot{\alpha} \ne 0$, then $\alpha' = 0$ for all $Y^a_{(A)} \in T_p \Sigma$. Furthermore, if there is any $Y^a_{(A)} \in T_p \Sigma$ such that $A' \ne 0$ then $\dot{A} = 0$ for all $\xi^a_{(i)} \in T_p N$, which reduces to the case (a). Otherwise, we have $A' = 0$ for all $Y^a_{(A)} \in T_p \Sigma$.

\item[(d)] $\xi^a_{(i)} \in T_p N$ such that $\dot{\alpha} = 0$ and $\dot{A} \ne 0$, then $A' = 0$ for all $Y^a_{(A)} \in T_p \Sigma$. Furthermore, if there is any $Y^a_{(A)} \in T_p \Sigma$ such that $\alpha' \ne 0$ then $\dot{\alpha} = 0$ for all $\xi^a_{(i)} \in T_p N$, which reduces to the case (b). Otherwise, we have $\alpha' = 0$ for all $Y^a_{(A)} \in T_p \Sigma$.
\end{itemize}

If none of the conditions from above are met, then we are left with the general ``synchronized case'': For any $\xi^a_{(i)} \in T_p N$ either both $\dot{A}$ and $\dot{\alpha}$ vanish or neither does, and for any $Y^a_{(A)} \in T_p \Sigma$ either both $A'$ and $\alpha'$ vanish or neither does. Let us look more closely at the particular subcases of the symmetry noninheritance that have been described above.

\bigskip

\emph{Symmetry inheriting amplitude}. On any open subset of $Q$ where the conditions from the case (a) are met we have
$$A = A(y^1,\dots,y^{D-n}) \quad \textrm{and} \quad \alpha = \alpha(z^1,\dots,z^n) \ .$$
Using the theorem 5 from \cite{ISm15} we can narrow down the form of the phase $\alpha$ even further. 

\btm
Let the spacetime $(M,g_{ab},\phi)$ be an o-t symmetric solution of (\ref{eq:EOM}) of order $n$ with the energy-momentum tensor (\ref{eq:Tcmplx}), such that the complex scalar field $\phi$ has symmetry inheriting amplitude $A$. Then the most general form of the symmetry noninheriting phase is
\be
\alpha = c_0 + \sum_{i=1}^n c_i z^i \ ,
\ee
where $\{c_0,c_1,\dots,c_n\}$ are some real constants.
\etm

\medskip

As the constant part of the phase, given by $c_0$, is irrelevant, we may always choose $c_0 = 0$. The rest of the constants $c_i$ may be constrained in the presence of a black hole. If the spacetime contains a Killing horizon $H[\chi]$ as defined in \ref{def:assKH}, then (\ref{eq:Tchichi}) implies that
\be
\sum_{i=1}^n b_i \Lie_{\xi_{(i)}} \alpha = 0 \ .
\ee
More concretely, if the Killing vector $\chi^a$ is given by (\ref{eq:chikm}), then the following relation holds
\be\label{eq:sumalpha}
\Lie_k \alpha + \sum_{i=1}^\ell \Omega_i \Lie_{m_{(i)}} \alpha = 0 \ .
\ee
This equation generalizes the ``resonance condition'' \cite{HKRS15} (see also \cite{BHR14}) for the hairy five-dimensional Myers-Perry black holes (MPBHsSH). As it represents the threshold between the decaying and the superradiant regimes, this was recognized \cite{HR14,HR15} as a hint for the existence of the new type of the black hole scalar hair.

\medskip

One might expect from the relation (\ref{eq:sumalpha}) that the HR solution \cite{HR14,HR15} has no symmetry noninheriting static limit $\Omega_i \to 0$, since this would imply $\Lie_k \alpha = 0$. However, we may still have the angular dependence of the phase ($\Lie_{m_{(i)}} \alpha \ne 0$ for some $m_{(i)}^a$) in a static axially symmetric spacetime with the nonrotating Killing horizon $H[k]$, see e.g.~\cite{CPS10,KKRS16}.

\bigskip

\emph{Symmetry inheriting phase}. On any open subset of $Q$ where the conditions from the case (b) are met we have
$$A = A(z^1,\dots,z^n) \quad \textrm{and} \quad \alpha = \alpha(y^1,\dots,y^{D-n}) \ .$$
Furthermore, from the fact that $\xi_{(i)}^a$ and $Y_{(A)}^a$ commute, it follows that
\be
\Lie_{\xi_{(i)}} E(Y_{(A)},Y_{(A)}) = 0 \ .
\ee 
Using this in the equation of motion (\ref{eq:EOM}) and the energy-momentum tensor (\ref{eq:Tcmplx}) we have
\be
\left( \alpha'^2 + \frac{\mathscr{Y}_A}{D-2}\,\frac{\dd V}{\dd(A^2)} \right) A \dot{A} = 0 \ ,
\ee
where $\mathscr{Y}_A \equiv g_{ab} Y_{(A)}^a Y_{(A)}^b$. So, if there is at least one $\xi^a_{(i)} \in T_p N$ such that $\dot{A} \ne 0$ (otherwise we would have the complete symmetry inheritance), then the expression in the parenthesis must vanish. 

\medskip

Here we have two possibilities: Either all $Y_{(A)}^a$ are null ($\mathscr{Y}_A = 0$), in which case the phase $\alpha$ is necessarily a constant and the energy-momentum tensor (\ref{eq:Tcmplx}) takes the form of a real scalar field (which was analysed in the previous section), or there is at least one non-null $Y_{(A)}^a$, in which case we can express $\dd V/\dd(A^2)$ with $\alpha'$ and $\mathscr{Y}_A$, both symmetry inheriting quantities (invariant with respect to all $\xi^a_{(i)}$). In the latter case $\dd V/\dd(A^2)$ has to be constant, which means that under given assumptions we necessarily have the simplest, mass term potential $V = \mu^2 A^2$! Detailed analysis of the symmetry inheritance of the complex scalar field with such a potential has been perfomed in \cite{ISm15}. In order to have the symmetry noninheriting amplitude, $\dot{A} \ne 0$, which is either bounded or periodic function, the associate Killing vector $\xi^a_{(i)}$ has to be a spacelike, hypersurface orthogonal and its norm $g_{ab} \xi^a_{(i)} \xi^b_{(i)}$ a constant. 

\medskip

If any such a solution exists, it would need to pass through a very narrow window left by the constraints obtained above. For example, suppose we have a static spacetime with the associated hypersurface-orthogonal Killing vector field $k^a$ (timelike at infinity), containing the associated Killing horizon $H[k]$. Is it possible to have an ergoregion where $k^a$ becomes spacelike, allowing at least in principle the existence of the complex scalar field with the time-dependent amplitude and the time-independent phase? The Vishveshwara-Carter's theorem \cite{Vish68,Carter69} (see also \cite{CostaPhD} for a technically polished version of the proof) guarantees that ergosurfaces, consisting of points where $k^a$ becomes null vector field, coincide with the Killing horizon $H[k]$, i.e.~there are no ergoregions in the domain of outer communications $\left<\left< M_{\mathrm{ext}} \right>\right>$ (for a definition see \cite{HCC}). Therefore, there are no such a complex scalar field in a static spacetime with the nonrotating black hole.

\bigskip

\emph{Strictly sni fields}. On any open subset $S \subseteq Q$ where the conditions from the case (c) or (d) are met (which don't reduce to the previous cases (a) or (b)), we have
$$A = A(z^1,\dots,z^n) \quad \textrm{and} \quad \alpha = \alpha(z^1,\dots,z^n) \ .$$
Again, using the fact that $\xi_{(i)}^a$ and $Y_{(A)}^a$ commute, it follows that
\be
0 = \Lie_{\xi_{(i)}} T(Y_{(A)})_a = \frac{2A\dot{A}}{D-2}\,\frac{\dd V}{\dd(A^2)} \, Y_{(A)a} \ .
\ee
Note that at each point $p \in S$ where $Y_{(A)}^a \ne 0$ and $\dd V/\dd(A^2) \ne 0$, we have $\dot{A} = 0$ for all $\xi^a_{(i)} \in T_p N$ (leading again to the case (a)). As the zeros of the vector fields $Y_{(A)}^a$ are typically just a zero measure set, the only possibly interesting subcase here is that of the constant potential $V$ (e.g.~the massless complex scalar field with $V = 0$). Assuming that $V$ is a constant, for any triple $\xi^a_{(i)}$, $\xi^a_{(j)}$ and $\xi^a_{(k)}$ of Killing vector fields we have
\be
0 = \Lie_{\xi_{(k)}} T(\xi_{(i)},\xi_{(j)}) = \Lie_{\xi_{(k)}} C_{ij}
\ee
where we have introduced an abbreviation
\be
C_{ij} \equiv (\Lie_{\xi_{(i)}} A)(\Lie_{\xi_{(j)}} A) + A^2 (\Lie_{\xi_{(i)}} \alpha)(\Lie_{\xi_{(j)}} \alpha) \ .
\ee
As under the given assumptions we also have $C'_{ij} = 0$ for any $Y^a_{(A)}$, it follows that all $C_{ij}$ are in fact constants. Furthermore, in the presence of the Killing horizon $H[\chi]$ introduced in \ref{def:assKH}, using the condition (\ref{eq:Tchichi}), we may deduce that another constraint,
\be\label{eq:bbC}
\sum_{i=1}^n \sum_{j=1}^n b_i b_j C_{ij} = 0
\ee
holds on any subset of $S$ which intersects the horizon $H[\chi]$. If we introduce a $n \times n$ symmetric matrix $\bm{\mathsf{B}}$, with entries $\bm{\mathsf{B}}_{ij} = b_i b_j$, and two column matrices,
$$\bm{\mathsf{u}} = \left( \Lie_{\xi_{(1)}} A \, \cdots \, \Lie_{\xi_{(n)}} A \right)^{\!\mathsf{T}} \, \textrm{and} \,\ \bm{\mathsf{v}} = \left( \Lie_{\xi_{(1)}} \alpha \, \cdots \, \Lie_{\xi_{(n)}} \alpha \right)^{\!\mathsf{T}} ,$$
then the Eq.~(\ref{eq:bbC}) can be rewritten in the matrix form as
\be\label{eq:uBuvBv}
\bm{\mathsf{u}}^\mathsf{T} \bm{\mathsf{B}} \bm{\mathsf{u}} +  A^2\,\bm{\mathsf{v}}^\mathsf{T} \bm{\mathsf{B}} \bm{\mathsf{v}} = 0 \ .
\ee
The matrix $\bm{\mathsf{B}}$ has two distinct eigenvalues, zero (with multiplicity $n-1$) and $b_1^2 + \cdots + b_n^2$ (with multiplicity $1$), hence it is positive semidefinite matrix. Thus, (\ref{eq:uBuvBv}) implies that
\be
\bm{\mathsf{u}}^\mathsf{T} \bm{\mathsf{B}} \bm{\mathsf{u}} = 0 = \bm{\mathsf{v}}^\mathsf{T} \bm{\mathsf{B}} \bm{\mathsf{v}} \ ,
\ee
and from here follows that both $\bm{\mathsf{B}} \bm{\mathsf{u}}$ and $\bm{\mathsf{B}} \bm{\mathsf{v}}$ are zero matrices. In other words, 
\be
\Lie_\chi A = 0 = \Lie_\chi \alpha \ .
\ee
For example, if we have a static spacetime with the nonrotating Killing horizon $H[k]$, then it follows that both the amplitude and the phase have to inherit the stationary symmetry, $\Lie_k A = 0 = \Lie_k \alpha$. However, in the rotating case (\ref{eq:chikm}), these constraints are not sufficient to exclude the existence of the strictly sni scalar black hole hair.

\bigskip

\emph{General sni field with synchronised $A$ and $\alpha$}. If the field $\phi$ doesn't belong to any of the (a)--(d) cases, it is difficult to give some more concrete conclusions about the symmetry inheritance and the existence of the associated black hole hair.

\medskip

\section{Final remarks}

We have shown that at least for the restricted (yet abundant) orthogonal-transitive class of the gravitational field equations and associated symmetric spacetimes, one can provide a detailed classification of the symmetry noninheriting scalar fields. Building upon this, we have explored the extent of the no-hair theorems in the domain of sni scalar fields. While the real scalar sni black hole hair can be pretty much excluded, in the complex case the form of such a hair is heavily constrained. An important open question is whether a black hole can support a complex scalar hair which breaks the symmetries in both the amplitude and the phase. On the formal side, it would be interesting to find other examples of the gravitational field equations belonging to some of the o-t classes, apart from those mentioned in this paper, as well as the nonexamples, apart from the gravitational Chern-Simons terms. Finally, as the physically more realistic spacetimes will possess only the approximate symmetries, definition of which undoubtedly brings in a certain amount of ambiguity, it is important to find the rules of the approximate symmetry inheritance (a recent progress in this direction has appeared in \cite{KG15}).

\begin{acknowledgments}
This research has been supported by the Croatian Science Foundation under the project No.~8946.
\end{acknowledgments}

\bibliography{snih}

\end{document}